\documentclass[12pt]{article}

\usepackage[dvips,letterpaper]{geometry}
\usepackage{graphicx,psfrag,epsf}
\usepackage{times}
\usepackage{fullpage}
\usepackage{setspace}
\usepackage{amsmath,amssymb,amsfonts}
\usepackage{bm}
\usepackage{graphicx}
\usepackage{epsfig}
\usepackage{subfigure}
\usepackage{url}
\usepackage{verbatim}
\usepackage{natbib}
\usepackage{color}
\usepackage{booktabs}
\usepackage{graphicx,psfrag,epsf}
\usepackage{algorithm,algpseudocode}
\usepackage{listings}
\usepackage{tikz}

\usepackage{graphicx}
\usepackage{amsmath, amssymb, amsfonts}
\usepackage{mathtools, here}
\usepackage{breakcites}
\usepackage{bm}
\usepackage{multirow}
\usepackage{subfigure}
\usepackage{url}
\usepackage{verbatim}
\usepackage{blkarray}
\usepackage{color}
\usepackage{booktabs}
\usepackage{physics}
\usepackage{lineno}
\usepackage{listings}
\usepackage{tikz}
\usepackage{ulem}

\input cyracc.def

\def \build#1#2#3{\mathrel{\mathop{#1}\limits^{#2}_{#3}}}

\def\diag{\mathop{\rm diag}\nolimits}

\def \build#1#2#3{\mathrel{\mathop{#1}\limits^{#2}_{#3}}}

\def \p {\partial}

\title{\bf {\Large {Improved estimators in beta prime regression models}}}

\author{
\textbf{\normalsize Francisco M.C. Medeiros}$^{1}$,\, \textbf{\normalsize Mariana C. Ara\'ujo}$^{1}$ \textbf{\normalsize and } \textbf{\normalsize Marcelo Bourguignon}$^{1}$\\[-0.1cm]
{\small $^{1}$Department of Statistics, Universidade Federal do Rio Grande do Norte, Brazil}\\[-0.15cm]
}

\date{}

\begin{document}

\maketitle

\begin{abstract}
\noindent In this paper, we consider the beta prime regression model recently proposed by \cite{bour18}, which is tailored
to situations where the response is continuous and restricted to the positive real line with skewed and long tails
and the regression structure involves regressors and unknown parameters.
We consider two different strategies of bias correction of the maximum-likelihood estimators
for the parameters that index the model. In particular, we discuss bias-corrected estimators for the mean and the dispersion
parameters of the model. Furthermore, as an alternative to the two analytically bias-corrected estimators discussed,
we consider a bias correction mechanism based on the parametric bootstrap.
The numerical results show that the bias correction
scheme yields nearly unbiased estimates.
An example with real data is presented and discussed.

\end{abstract}

\section{Introduction}
\noindent

The beta prime (BP) distribution  (known as inverted beta distribution or beta distribution of the second kind as well) is a two-parameter distribution on the positive real line, which can be interpreted as the distribution of the odds ratio of a variable distributed according to the beta distribution, i.e., if $X$ has a beta distribution with parameters $\alpha$ and $\beta$, then $Y = X/(1-X)$ has a BP distribution with $\alpha > 0$ and $\beta > 0$ both shape parameters. We are adopting the parameterization for the BP distribution in terms of the mean and precision parameters which was proposed by \cite{bour18}. An advantage of using this parameterization is that we can introduce regression structures for each mean and precision parameters and the interpretation of the regression coefficients is straightforward in terms of them as in generalized linear models. Thus, the BP random variable $Y$ (\cite{bour18}) is defined as follows: Let $Y$ be a random variable with probability density function (pdf) given by
\begin{equation}\label{inv:011}
f(y; \mu, \phi) = \frac{y^{\mu(\phi + 1)-1}(1 + y)^{-[\mu(\phi + 1) + \phi + 2]}}{B(\mu(1+ \phi), \phi + 2)}, \quad y > 0,
\end{equation}
where $\mu > 0$ and $\phi > 0$ are mean and precision parameters, respectively, $B(\mu(1+ \phi), \phi + 2) = \Gamma(\mu(1+ \phi))\Gamma(\phi + 2)/\Gamma(\mu(1+ \phi) + \phi + 2)$ is the beta function and $\Gamma(\mu(1+ \phi)) = \int_{0}^{\infty}\omega^{\mu(1+ \phi)-1}\textrm{e}^{-\omega}\textrm{d} \omega$ is the gamma function. From now on, we use the notation $Y \sim \textrm{BP}(\mu, \phi)$ to indicate that $Y$ is a random variable following a BP distribution. The mean and variance of $Y$ are
$$
\textrm{E}[Y] = \mu
\quad
\textrm{and}
\quad
\textrm{Var}[Y] = \frac{\mu(1+\mu)}{\phi}.
$$

Some features of the BP model are \citep{bour18}: first, the variance function of the BP model assumes a quadratic form similar to the gamma distribution. However, the variance function of the proposed model is larger than the variance function of gamma distribution, which may be more appropriate in certain practical situations; second, the BP hazard rate function can have an upside-down bathtub or increasing depending on the parameter values. The most classical two-parameter distributions such as Weibull and gamma distributions have monotone hazard rate functions; third, the skewness and kurtosis of the BP distribution can be much larger than those of the gamma and inverse gaussian distributions; fourth, there are some stochastic representation of the BP random variable.
%The $r$th moment about zero of $Y$ is given by
%\begin{equation*}\label{inv:04}
%\textrm{E}[Y^r] = \frac{B(\alpha + r, \beta - r)}{B(\alpha, \beta)}, \quad -\alpha < r < \beta.
%\end{equation*}

In the literature there are only a few works dealing with the BP distribution.
\cite{McDonald:1987aa} discussed its properties and obtained the maximum likelihood (ML) estimates of the model parameters.
Bias-corrected versions of the MLEs of the parameters that index the BP distribution were obtained by \cite{borkocordeiro}. It is worth mention that all the works related above have considered the usual parameterization of the BP distribution. Considering the parameterization we adopted, \cite{bour18} used the the ML method for estimating the parameters that index the BP regression model. However, as can be seen in Table 2 in \cite{bour18},  in small-sized samples, the ML estimators of these parameters (especially for precision structure) may be extremely biased.
So, it is important consider alternative estimators with smaller biases when the number of observations are small.

Investigates how the maximum likelihood estimator behaves in small-sized sample, in particular bias analysis, is an important research area. In regular parametric statistical models the maximum likelihood estimator bias is generally of the order $\mathcal{O}(n^{-1}) $ for large sample size $n$ and are, in practice, usually ignored since that the asymptotic standard error is of order $O(n^{-1/2})$. When dealing with small-sized sample, however, bias can be a problematic issue, thus it can not be neglected. So, it is important to obtain bias correction in these cases. Bias reduction was studied by several authors. In uniparametric models, \cite{Bartlett1953} obtained an expression for the $\mathcal{O}(n^{-1})$ bias from the maximum likelihood estimator. Assuming independent, but not necessarily identically distributed observations, \cite{coxsnell68} obtained a general expression for the $\mathcal{O}(n^{-1})$ bias of the maximum likelihood estimator in multiparametric models. This result has become widely used in the literature to obtain general expressions for the $\mathcal{O}(n^{-1})$ bias and to propose bias-corrected estimators in various parametric models. For instance, \cite{Lemonte-et-al-2007}, \cite{Cysneiros-et-al-2010}, \cite{Simas2011}, \cite{SouzaVasconcellos2011} and \cite{melo2018}. %\cite{Lamb2000}, \cite{SahaPaul2005}

Usually, the approach to obtain bias-corrected versions of the MLEs uses the second order bias. In this procedure the adjustment is made after the MLEs were computed. Additionally, an alternative approach was proposed by \cite{Firth1993}, who suggested that a bias reduction method by modifying the score function previous to obtain the parameter estimates. This method is called the preventive method and has been studied in parametric models where maximum likelihood estimates can be unstable (infinite or belonging to the parametric space boundary) such as \cite{Bull2002}, \cite{Sartori2006}, \cite{Kosmidis2009}, \cite{Kosmidis2011} and \cite{Kosmidis2014}. Another possible way to perform bias correction is through bootstrap resampling, which requires no explicit derivation of the bias function.
In this context, the main goal of this paper is to derive a closed-form expression for the second order biases of the ML estimators in the BP regression model which can be used to define bias corrected ML estimators to order $\mathcal{O}(n^{-1})$.

This paper is organized as follows: this introductory
section. In Section \ref{sec2}, the BP regression model is introduced and some of
its basic properties are outlined.
In Section \ref{sec3}, we obtain the second order biases of the MLEs of the means of the responses and precision parameters of the model.
Section \ref{sec4} discusses the numerical results.
In Section \ref{sec5}, we consider an empirical example. Finally, Section \ref{sec6} concludes the paper.

\section{Beta prime regression model}\label{sec2}
\noindent

Consider $n$ independent random variables $Y_1, \ldots, Y_n$ where each $Y_i$, $i = 1, \ldots, n$ has BP distribution with pdf given by \eqref{inv:011} with mean $\mu_i$ and precision parameter $\phi_i$. \cite{bour18} proposed the BP regression model which is defined by \eqref{inv:011} and by two functional relations
\begin{equation}\label{cs1}
g_1(\mu_i) = \eta_{1i} = \mathbf{x}^\top_i\bm{\beta} \quad \textrm{and} \quad g_2(\phi_i) = \eta_{2i} = \mathbf{z}^\top_i\bm{\nu},
\end{equation}
where $g_1: \mathbb{R} \rightarrow \mathbb{R}^+$ and $g_2: \mathbb{R} \rightarrow \mathbb{R}^+$ are strictly monotone, positive and at least twice differentiable link functions, $\eta_{1i}$ and $\eta_{2i}$ are the linear predictors, $\bm{\beta} = (\beta_1, \ldots, \beta_p)^\top$ $(\bm{\beta} \in \mathbb{R}^p)$ and $\bm{\nu} = (\nu_1, \ldots, \nu_q)^\top$ ($\bm{\nu} \in \mathbb{R}^q, \ \ q< n-p$) are unknown parameter vectors to be estimated, and $\mathbf{x}_i = (x_{i1}, \ldots, x_{ip})^\top$ and $\mathbf{z}_i = (z_{i1}, \ldots, z_{iq})^\top$ are observations on $p$ and $q$ known regressors, for $i = 1, \ldots, n$. Additionally, we assume that the covariate matrices $\mathbf{X} = (\mathbf{x}_1, \ldots, \mathbf{x}_n)^\top$ and $\mathbf{Z} = (\mathbf{z}_1, \ldots, \mathbf{z}_n)^\top$ have rank $p$ and $q$, respectively. Besides the interpretation of the regression coefficients being in terms of the mean and precision parameters, another advantage of the model proposed by (\ref{inv:011}) and (\ref{cs1}) is that it is suitable for modeling asymmetric data, being an alternative to the generalized linear models when dealing with asymmetric dataset.

The log-likelihood function for $(\bm{\beta}, \bm{\nu})$ given the observed values $y_1, \ldots, y_n$ is
\begin{eqnarray}\label{logm}
\ell(\bm{\beta}, \bm{\nu}) = \sum_{i=1}^{n}\ell(\mu_i, \phi_i),
\end{eqnarray}
being
\begin{eqnarray*}
\ell(\mu_i, \phi_i)&=& [\mu_i(1+\phi_i) - 1]\log(y_i) - [\mu_i(1+ \phi_i) + \phi_i +2]\log(1+y_i) \nonumber \\
&-&\log[ \Gamma(\mu_i(1+\phi_i))]- \log[ \Gamma(\phi_i+2)] + \log[ \Gamma(\mu_i(1+\phi_i) + \phi_i +2)];
\end{eqnarray*}
$\mu_i=g_1^{-1}(\eta_{1i})$ and $\phi_i=g_2^{-1}(\eta_{2i})$ are functions of $\bm{\beta}$ and $\bm{\nu}$, respectively, as defined in (\ref{cs1}). A method for obtaining the parameters estimates of the BP model defined by (\ref{inv:011}) and (\ref{cs1}) is described in details in \cite{bour18}. They consider the \textit{gamlss} function for this purpose.

We assume that the log-likelihood function (\ref{logm}) satisfies the usual regularity conditions of large sample likelihood theory \citep[see][]{cookwei:83}. Thus, when $n$ is large and under some regular conditions we have that
\[
\left(\begin{array}{c}
\bm{\widehat{\beta}}\\
{\bm{\widehat{\nu}}}
\end{array}\right)
\build{\sim}{a}{} \mbox{N}_{p + q}\left(
\left(\begin{array}{c}
\bm{{\beta}}\\
{\bm{{\nu}}}
\end{array}\right)
, \mathbf{K}(\bm{\widehat{\beta}},{\bm{\widehat{\nu}}})^{-1}\right),
\]
where $\build{\sim}{a}{}$ means ``approximately distributed'' and $\mathbf{K}(\bm{\widehat{\beta}},{\bm{\widehat{\nu}}})^{-1}$ is the inverse of Fisher's information matrix evaluated at $\bm{\widehat{\beta}}$ and ${\bm{\widehat{\nu}}}$, which can be approximated by $J(\bm{\widehat{\beta}},{\bm{\widehat{\nu}}})^{-1}$, where $-J$ denotes the $(p+q) \times (p+q)$ Hessian matrix evaluated at $(\bm{\widehat{\beta}}^\top,{\bm{\widehat{\nu}}}^\top)^\top$. Fisher's information matrix $\mathbf{K}(\bm{\beta},{\bm{\nu}})$ is presented in Appendix A.

\section{Bias correction of the MLEs}\label{sec3}
\noindent

Let $\bm{\theta}=(\bm{\beta}^\top, \bm{\nu}^\top)^\top$ be the unknown parameter vector of the BP regression model. We now obtain an expression for the second order biases of the MLEs of the components of $\bm{\theta}$ using Cox and Snell's (\cite{coxsnell68}) general formula. In order to obtain this expression, we first introduce some notation. The lower subscripts $r, s, t, u, \ldots$ and the upper subscripts $R, S, T, U, \ldots$ denote, respectively, the components of $\bm{\beta}$ and $\bm{\nu}$ vectors. Therefore, the partional derivatives of the log-likelihood (\ref{logm}) with respect to the components of $\bm{\beta}$ and $\bm{\nu}$ are presented as  $U_{r}= \partial \ell/\partial \beta_r,$ $U_{rS}= \partial^2 \ell/ \partial \beta_r \partial \nu_S,$ $U_{rST}=\partial^3 \ell/ \partial \beta_r \partial\nu_S \partial \nu_T,$ etc. The moments of the log-likelihood derivatives are represented by $\kappa_{rs}=E(U_{rs}),$ $\kappa_{r,s}=E(U_r U_s),$ $\kappa_{r, ST}=E(U_r U_{ST}),$ etc, where all $\kappa's$ regard to a total covering the whole sample and are, in general, of order $\mathcal{O}(n^{-1})$. The moments derivatives are defined by $\kappa_{rs}^{(t)}= \partial \kappa_{rs}/\partial \beta_t,$ $\kappa_{rs}^{(T)}= \partial \kappa_{rs}/\partial \nu_T,$ etc. Finally, we denote the elements of the inverse of Fisher's information matrix $K(\bm{\beta}, \bm{\nu})^{-1}=K(\bm{\theta})^{-1},$ which are $\mathcal{O}(n^{-1})$, as $\kappa^{r,s} = -\kappa^{rs},$ $\kappa^{r,S} = -\kappa^{rS},$ $\kappa^{R,s} = -\kappa^{Rs}$ and $\kappa^{R,S} = -\kappa^{RS}.$

From the general Cox and Snell's (\citeyear{coxsnell68}) formula we can obtain the $\mathcal{O}(n^{-1})$ bias of the MLE for the $a$th component of the parameter vector $\bm{\widehat{\theta}}=(\widehat{\theta}_1, \ldots, \widehat{\theta}_p,\\ \widehat{\theta}_{p+1}, \ldots, \widehat{\theta}_{p+q})^\top = (\bm{\widehat{\beta}}^\top, \bm{\widehat{\nu}}^\top)^\top$ as:

{\footnotesize \begin{eqnarray}\label{vcs}
B_{\widehat{\bm{\theta}}}(\bm{\theta}_a) &=&\sum_{r,s,u}\kappa^{ar}\kappa^{su} \left \{ \kappa_{rs}^{(u)} - \frac{1}{2}\kappa_{rsu} \right \} + \sum_{R, s, u} \kappa^{aR} \kappa^{su} \left \{ \kappa_{Rs}^{(u)} - \frac{1}{2} \kappa_{Rsu}\right \} \nonumber \\
&+& \sum_{r, S, u} \kappa^{ar} \kappa^{Su} \left \{ \kappa_{rS}^{(u)} - \frac{1}{2} \kappa_{rSu}\right \} + \sum_{r,s,U}\kappa^{ar}\kappa^{sU} \left \{ \kappa_{rs}^{(U)} - \frac{1}{2}\kappa_{rsU} \right \} \nonumber \\
 &+& \sum_{R, S, u} \kappa^{aR} \kappa^{Su} \left \{ \kappa_{RS}^{(u)} - \frac{1}{2} \kappa_{RSu}\right \} + \sum_{R, s, U} \kappa^{aR} \kappa^{sU} \left \{ \kappa_{Rs}^{(U)} - \frac{1}{2} \kappa_{RsU}\right \} \nonumber \\
&+&\sum_{r,S,U}\kappa^{ar}\kappa^{SU} \left \{ \kappa_{rS}^{(U)} - \frac{1}{2}\kappa_{rSU} \right \} + \sum_{R, S, U} \kappa^{aR} \kappa^{SU} \left \{ \kappa_{RS}^{(U)} - \frac{1}{2} \kappa_{RSU}\right \}.
\end{eqnarray} }

From (\ref{eqm22}), we can observe that $\bm{\beta}$ and $\bm{\nu}$ are not orthogonal, hence all terms in (\ref{vcs}) must be considered. In order to save space, all cumulants needed to obtain (\ref{vcs}) are given in Appendix B. After a long algebra presented in details in Appendix B we achieve to the expressions for the second order biases of $\bm{\widehat{\beta}}$ and $\bm{\widehat{\nu}},$ given in matrix form respectively by
\begin{eqnarray*}
B_{\bm{\widehat{\beta}}}(\bm{\beta}) & = & K^{\bm{\beta \beta}} X^\top \left[ M_1 P_{\bm{\beta \beta}} + (M_2 + M_3) P_{\bm{\beta \nu}} + M_5 P_{\bm{\nu \nu}} \right] \\
										& + & K^{\bm{\beta \nu}} Z^\top \left[ M_2 P_{\bm{\beta \beta}} + (M_4 + M_5)P_{\bm{\beta \nu}} + M_6 P_{\bm{\nu \nu}} \right]
\end{eqnarray*}
and
\begin{eqnarray*}
B_{\bm{\widehat{\nu}}}(\bm{\nu}) & = & K^{\bm{\beta \nu}} X^\top \left[M_1P_{\bm{\beta \beta}} + (M_2 + M_3)P_{\bm{\beta \nu}} + M_5 P_{\bm{\nu \nu}} \right]\\
									& + & K^{\bm{\nu \nu}} Z^\top \left[ M_2P_{\bm{\beta \beta}} + (M_4 + M_5)P_{\bm{\beta \nu}} + M_6 P_{\bm{\nu \nu}} \right],
\end{eqnarray*}
where $K^{\bm{\beta \beta}},$ $K^{\bm{\beta \nu}}$ and $K^{\bm{\nu \nu}}$ represent matrices which components are respectively the $(r,s)$th, $(r,S)$th and $(R,S)$th elements of the inverse of Fisher's information matrix, $M_1$ to $M_6$ are presented in Appendix B, $P_{\bm{\beta \beta}},$ $P_{\bm{\beta \nu}}$ and $P_{\bm{\nu \nu}}$ are vectors with the same $n \times 1$ dimension and which elements are the diagonal elements of $X K^{\bm{\beta \beta}} X^\top,$ $X K^{\bm{\beta \nu}} Z^\top$ and $Z K^{\bm{\nu \nu}} Z^\top,$ respectively.

We now assume the $2n \times 1$ vector $\delta_1$ defined as
\begin{equation*}
\delta_1 = \left(\begin{array}{cc}
     M_1P_{\bm{\beta \beta}} + (M_2 + M_3)P_{\bm{\beta \nu}} + M_5 P_{\bm{\nu \nu}} \\
     M_2P_{\bm{\beta \beta}} + (M_4 + M_5)P_{\bm{\beta \nu}} + M_6 P_{\bm{\nu \nu}}
\end{array} \right),
\end{equation*} and consider $K^{\bm{\beta *}}=(K^{\bm{\beta \beta}}\,\,K^{\bm{\beta \nu}})$ and $K^{\bm{\nu *}}=(K^{\bm{\nu \beta}} \,\, K^{\bm{\nu \nu}}),$ the $p \times (p+q)$ upper and $q \times (p+q)$ lower blocks of the matrix $K(\bm{\beta}, \bm{\nu})^{-1},$ respectively. Thus, we can express the second-order biases of $\widehat{\bm{\beta}}$ and $\widehat{\bm{\nu}}$ as
$$B_{\bm{\widehat{\beta}}}(\bm{\beta}) = K^{\bm{\beta *}} \mathbb{X}^\top \delta_1 \ \ \ \mbox{and} \ \ \ B_{\widehat{\bm{\nu}}}(\bm{\nu}) = K^{\bm{\nu *}} \mathbb{X}^\top \delta_1.$$
From the expressions above, we can obtain in matrix form the second order bias of the MLE of the joint vector $\bm{\theta}=(\bm{\beta}^\top, \bm{\nu}^\top)^\top$ expressed as
$$B_{\widehat{\bm{\theta}}}(\bm{\theta})=(\mathbb{X}^\top \widetilde{\bm{K}} \mathbb{X})^{-1}\mathbb{X}^\top \delta_1.$$

Now, we define the bias-corrected estimator as
$$\widetilde{\bm{\theta}}= \widehat{\bm{\theta}} - B_{\widehat{\bm{\theta}}}(\widehat{\bm{\theta}}),$$
%\[
%\widetilde{\bm{\beta}} = \widehat{\bm{\beta}} - %B_{\widehat{\bm{\beta}}}(\widehat{\bm{\beta}}) \hspace{0.5cm} \mathrm{and} %\hspace{0.5cm} \widetilde{\bm{\nu}} = \widehat{\bm{\nu}} - %B_{\widehat{\bm{\nu}}}(\widehat{\bm{\nu}}),
%\]
%where $B_{\widehat{\bm{\beta}}}(\widehat{\bm{\beta}})$ and $B_{\widehat{\bm{\nu}}}(\widehat{\bm{\nu}})$ are biases of the $\widehat{\bm{\beta}}$ and $\widehat{\bm{\nu}}$ with the unknown parameters replaced by their MLEs.
where $B_{\widehat{\bm{\theta}}}(\widehat{\bm{\theta}})$ is bias of the $\widehat{\bm{\theta}}$ with the unknown parameters replaced by their MLEs. Considering the assumptions assumed in Section \ref{sec2}, we have that the asymptotic distribution of $\bm{\theta}$ is $\mbox{N}_{p + q}(\bm{\theta}, \mathbf{J}(\bm{\theta})^{-1} ),$ where $\mathbf{J}(\bm{\theta}) = \mathbf{J}(\bm{\beta},{\bm{\nu}})^{-1}$
%\[
%\left[\begin{array}{c}
%\bm{\widetilde{\beta}}\\
%{\bm{\widetilde{\nu}}}
%\end{array}\right]
%
%\ \ \mbox{is} \ \
%
%\mbox{N}_{p + q}\left(
%\left[\begin{array}{c}
%\bm{{\beta}}\\
%{\bm{{\nu}}}
%\end{array}\right]
%, \mathbf{J}(\bm{\beta},{\bm{\nu}})^{-1}\right).
%\]

A second approach to correct the second order bias of the MLE of $\bm{\theta}=(\bm{\beta}^\top, \bm{\nu}^\top)^\top$ is considering the ``preventive'' method proposed by \cite{Firth1993}. This method basically consists of modify the original score function in order to remove the $\mathcal{O}(n^{-1})$ bias. The modified score function is given by
\[
U^*(\bm{\theta}) = U(\bm{\theta}) -
K(\bm{\theta})B_{\widehat{\bm{\theta}}}(\bm{\theta}),
\]
being $K(\bm{\theta})$ the information matrix and $B_{\widehat{\bm{\theta}}}(\bm{\theta})$ the $\mathcal{O}(n^{-1})$ bias. Considering the BP regression model and replacing the expression obtained for $B_{\widehat{\bm{\theta}}}(\bm{\theta}),$ the modified score function has the following form:$$U^*(\bm{\theta}) = U(\bm{\theta}) -  \mathbb{X}^\top \delta_1.$$ The second order bias corrected MLEs $\check{\bm{\theta}}$ is the solution of $U^*(\bm{\theta})=\bm{0}$. Also, $\check{\bm{\theta}}$ is asymptotically normal distributed as $\mbox{N}_{p + q}(\bm{\theta}, \mathbf{J}(\bm{\theta})^{-1} ),$ with $\mathbf{J}(\bm{\theta})$ as given previously.

Another way to bias-correcting the MLEs of the regression parameters is by the bootstrap technique \citep[see, for example,][]{EfronTibshirani1993}. In this paper, in order to reduce the computational burden, we shall adopt the warp-speed bootstrap method of \citep{Giacomini-et-al-2013} for evaluating the proposed resampling scheme. The warp-speed bootstrap method follows the steps described below. Instead of computing the MLEs for each Monte Carlo sample $r=1,2,\ldots,m$ (with $m$ being the total number of Monte Carlo replications) on the basis of $B$ bootstrap samples, just one resample (i.e.~$B=1$) is generated from the assumed model with the parameters replaced by estimates of maximum likelihood computed using the original sample for each Monte Carlo sample and, hence, estimates of maximum likelihood, say $\widehat{\bm{\theta}}^{*}$, is computed for that sample. Therefore, the bootstrap bias estimates $\widehat{\bm{\theta}}$ is
$$
B_{\widehat{\bm{\theta}}}(\widehat{\bm{\theta}}^{*}) = \widehat{\bm{\theta}}^{*} - \widehat{\bm{\theta}}.
$$
By using the bootstrap bias estimate presented above, we arrive at the following bias-corrected, to order ${\cal{O}}(n^{-1})$, estimator:
$$
\tilde{\bm{\theta}}^{b} = 2\widehat{\bm{\theta}} - \widehat{\bm{\theta}}^{*}.
$$

For a good discussion to the bootstrap method, see \citet[][Chapter~16]{EfronTibshirani1993}. Finally, it is worth mentioning that the idea behind the warp-speed bootstrap method is that taking just {\it one} bootstrap draw for each simulated sample is sufficient to provide a useful approximation to the bias of estimator. Applying this insight to Monte Carlo evaluation of bootstrap-based bias yields evaluation methods that work with $M=1$ \citep{Giacomini-et-al-2013}. Due to the resulting dramatic computational savings, \citep{Giacomini-et-al-2013} called their method as ``Warp-Speed'' Monte Carlo method. Therefore, the bootstrap-based bias on the basis of warp-speed bootstrap method become a viable alternative to inferential improvements in small samples when there are impeditive or too costly analytical difficulties.

\section{Numerical results}\label{sec4}
\noindent

We now present a Monte Carlo simulation study to investigate and compare the performance of the MLEs along with their corrected versions proposed in this article in small and moderate-sized samples. We use a BP regression models with dispersion covariates and a log link. We consider the model
\[
\log(\mu_i) = \beta_{0} + \sum_{\ell=1}^p \beta_{\ell}x_{i\ell}\quad \textrm{and} \quad \log(\phi_i) =\nu_{0} + \sum_{\ell=1}^q \nu_{\ell}x_{i\ell}, \quad i=1,2,\ldots,n,
\]
where the true values of the parameters were taken as 1. The covariates values are taken as random draws from the ${\cal {U}}(0,1)$ distribution and their values were held constant throughout the simulations. We consider different values for the number of regression parameters ($p$ and $q$) and the sample size ($n=30$, $40$ and $60$). The number of Monte Carlo replicates was $10.000$ and all the simulations were performed using the R language \citep{r:2017}. In each Monte Carlo replica, we computed the MLEs of the parameters, their corrected versions from the corrective method \citep{coxsnell68}, preventive method \citep{Firth1993}, and the parametric version of the bootstrap method \citep{Giacomini-et-al-2013}. In order to analyze the results, we computed, for each sample size and for each estimator, the mean of estimates, bias,
variance and mean square error (MSE). The results are presented in Tables \ref{tab1}, \ref{tab2} and \ref{tab3} for $p=q=1$, $p=q=2$, and $p=q=3$, respectively.

Tables \ref{tab1}-\ref{tab3} summarize the simulation results for the $\beta's$ and $\nu's$ varying the sample size $n$ and the number of regression parameters ($p$ and $q$). As can be seen in Tables \ref{tab1}-\ref{tab3}, for most part of the parameters the estimated biases, in absolute value, of the original MLEs were larger than the others. In general, for the $\beta's$, the biases of preventive estimators were smaller than those of the corrective estimators and bootstrap estimators. For the $\nu's$, the biases of the bootstrap estimators were, in general, smaller than those of the corrective estimators and the preventive estimators. These performances are independent of the number of parameters to be estimated. For instance, in absolute value, when $p=q=1$ and $n=30$ the bias of the parameter $\beta_0$ were $0.0048$ (MLE), $0.0009$ (Cox-Snell), $0.0000$ (Firth) and $0.0013$ (p-boot) and the bias of the parameter $\nu_0$ were 0.1044 (MLE), 0.0148 (Cox-Snell), 0.0083 (Firth) and 0.0064 (p-boot); see Table \ref{tab1}. However, for all parameters, in most cases the MSE of the corrective estimators were the smallest and the MSE of the bootstrap estimators were the largest, followed by the MSE of the preventive estimators. When we increase the sample size for the $\beta's$, the bootstrap estimators tends to shows the smallest bias, although they have the largest MSE. For instance, for $p = q = 3$ and $n = 30$  we have that the bias, in absolute value, for $\beta_1$ were $0.0026$ (MLE), $0.0024$ (Cox-Snell), $0.0016$ (Firth) and $0.0031$ (p-boot), while when $n=60$ they were $0.0018$ (MLE), $0.0016$ (Cox-Snell), $0.0015$ (Firth) and $0.0014$ (p-boot); see Table \ref{tab3}. Comparing the results presented in each Table, we can observe that as the sample size increases, in general, the bias of the estimators reduces, as expected.

The previous findings are confirmed by the box plots shown in Fig. 1, which were obtained for sample size $n = 30$.  In summary, the bias of the MLEs, especially for the $\nu's$ parameters, are larger than the bias of the corrected estimators. Box plots for different values of $n$, $p$, and $q$ (not shown) exhibited a similar pattern. Therefore, we recommend the use of method (Cox-Snell, Firth or parametric bootstrap) to reduce bias in small and moderate sample size.
%%%%%%%%%%%%%%%%%%%%%%%%%%%%%%%%%%%%%%%%%%%%%%%%%%%%%%%%%%%%%%%%%%%%%%%%%%%%%%
%              resultados p=q=1
%%%%%%%%%%%%%%%%%%%%%%%%%%%%%%%%%%%%%%%%%%%%%%%%%%%%%%%%%%%%%%%%%%%%%%%%%%%%%%
\begin{table}[!htb]
  \centering
\caption{Simulation results for $p=q=1$.}
%\rotatebox{90}{
\resizebox{\linewidth}{!}{
\begin{tabular}{lrrrrrrrrrrrrrr}
\hline
 & \multicolumn{4}{c}{$n=30$} && \multicolumn{4}{c}{$n=40$} && \multicolumn{4}{c}{$n=60$}\\
  \cline{2-5} \cline{7-10} \cline{12-15}
Estimates & MLE &Cox-Snell &Firth &p-boot && MLE &Cox-Snell & Firth & p-boot && MLE & Cox-Snell &Firth &p-boot \\ \hline
$\widehat{\beta_0}$ & 0.9952  & 0.9991  & 1.0000  & 0.9987  && 0.9952  & 0.9993  & 1.0000 & 0.9992 && 0.9965 & 0.9997 & 1.0001 & 0.9979 \\
Bias                & $-$0.0048 & $-$0.0009 & $-$0.0000 & $-$0.0013 && $-$0.0048 & $-$0.0007 & 0.0000 & $-$0.0008 && $-$0.0035 & $-$0.0003 & 0.0001 & $-$0.0021 \\
variance            & 0.0296  & 0.0297  & 0.0298  & 0.0540  && 0.0195  & 0.0196  & 0.0196 & 0.0363 && 0.0124 & 0.0124 & 0.0125 & 0.0237 \\
MSE                 & 0.0296  & 0.0297  & 0.0298  & 0.0540  && 0.0195  & 0.0196  & 0.0196 & 0.0363 && 0.0124 & 0.0124 & 0.0125 & 0.0237 \\
\\
$\widehat{\beta_1}$ & 0.9955  & 1.0005 & 1.0019 & 1.0010 && 0.9988  & 1.0004 & 1.0007 & 1.0005 && 1.0004 & 1.0007 & 1.0007 & 1.0033 \\
Bias                & $-$0.0045 & 0.0005 & 0.0019 & 0.0010 && $-$0.0012 & 0.0004 & 0.0007 & 0.0005 && 0.0004 & 0.0007 & 0.0007 & 0.0033 \\
variance            & 0.0827  & 0.0824 & 0.0824 & 0.1509 && 0.0569  & 0.0568 & 0.0567 & 0.1070 && 0.0339 & 0.0339 & 0.0340 & 0.0660 \\
MSE                 & 0.0827  & 0.0824 & 0.0825 & 0.1509 && 0.0569  & 0.0568 & 0.0567 & 0.1070 && 0.0339 & 0.0339 & 0.0340 & 0.0660 \\
\\

$\widehat{\nu_0}$  & 1.1044 & 1.0148 & 1.0083 & 1.0064 && 1.0872 & 1.0116 & 1.0094 & 1.0079 && 1.0624 & 1.0068 & 1.0045 & 1.0111 \\
Bias               & 0.1044 & 0.0148 & 0.0083 & 0.0064 && 0.0872 & 0.0116 & 0.0094 & 0.0079 && 0.0624 & 0.0068 & 0.0045 & 0.0111 \\
variance           & 0.6015 & 0.5083 & 0.6278 & 1.0849 && 0.3770 & 0.3412 & 0.3728 & 0.6987 && 0.2529 & 0.2372 & 0.2676 & 0.4702 \\
MSE                & 0.6124 & 0.5085 & 0.6278 & 1.0849 && 0.3847 & 0.3413 & 0.3729 & 0.6988 && 0.2568 & 0.2373 & 0.2676 & 0.4703 \\
\\
$\widehat{\nu_1}$  & 1.1215 & 1.0102 & 0.9939  & 1.0016 && 1.0699 & 1.0013 & 0.9913  & 0.9915 && 1.0320 & 0.9965 & 0.9941 & 0.9778 \\
Bias               & 0.1215 & 0.0102 & $-$0.0061 & 0.0016 && 0.0699 & 0.0013 & $-$0.0087 & $-$0.0085&& 0.0320 & $-$0.0035 & $-$0.0059 & $-$0.0222 \\
variance           & 1.7683 & 1.4470 & 1.8151  & 3.0861 && 0.9782 & 0.8591 & 0.9475  & 1.7650 && 0.6134 & 0.5640 & 0.6468 & 1.1392 \\
MSE                & 1.7831 & 1.4471 & 1.8152  & 3.0861 && 0.9831 & 0.8591 & 0.9475  & 1.7651 && 0.6144 & 0.5640 & 0.6468 & 1.1397 \\
\hline
\end{tabular}
}
%}
\label{tab1}
\end{table}

%%%%%%%%%%%%%%%%%%%%%%%%%%%%%%%%%%%%%%%%%%%%%%%%%%%%%%%%%%%%%%%%%%%%%%%%%%%%%%
%              resultados p=q=2
%%%%%%%%%%%%%%%%%%%%%%%%%%%%%%%%%%%%%%%%%%%%%%%%%%%%%%%%%%%%%%%%%%%%%%%%%%%%%%

\begin{table}[!htb]
  \centering
\caption{Simulation results for $p=q=2$.}
\resizebox{\linewidth}{!}{
\begin{tabular}{lrrrrrrrrrrrrrr}
\hline
 & \multicolumn{4}{c}{$n=30$} && \multicolumn{4}{c}{$n=40$} && \multicolumn{4}{c}{$n=60$}\\
  \cline{2-5} \cline{7-10} \cline{12-15}
Estimates & MLE &Cox-Snell &Firth &p-boot && MLE &Cox-Snell & Firth & p-boot && MLE & Cox-Snell &Firth &p-boot \\ \hline
$\widehat{\beta_0}$ & 0.9934  & 0.9966  & 0.9992  & 0.9981 && 0.9934 & 0.9976 & 0.9991 & 0.9999     && 0.9958 & 1.0003 & 1.0013 & 1.0006\\
Bias                & $-$0.0066 & $-$0.0034 & $-$0.0008 & $-$0.0019&& $-$0.0066 & $-$0.0024 & $-$0.0009 & $-$0.0001 && $-$0.0042 & 0.0003 & 0.0013 & 0.0006\\
variance            & 0.0273  & 0.0274  & 0.0286  & 0.0480 && 0.0205 & 0.0205 & 0.0206 & 0.0378     && 0.0130 & 0.0130 & 0.0130 & 0.0245\\
MSE                 & 0.0274  & 0.0274  & 0.0286  & 0.0480 && 0.0206 & 0.0205 & 0.0206 & 0.0378     && 0.0130 & 0.0130 & 0.0130 & 0.0245\\\\
$\widehat{\beta_1}$ & 0.9970  & 0.9995  & 0.9994  & 0.9989 && 1.0018 & 1.0023 & 1.0025 & 0.9995  && 1.0019 & 1.0020 & 1.0018 & 0.9992\\
Bias                & $-$0.0030 & $-$0.0005 & $-$0.0006 & $-$0.0011&& 0.0018 & 0.0023 & 0.0025 & $-$0.0005 && 0.0019 & 0.0020 & 0.0018 & $-$0.0008\\
variance            & 0.0494  & 0.0493  & 0.0508  & 0.0872 && 0.0356 & 0.0355 & 0.0356 & 0.0660  && 0.0225 & 0.0225 & 0.0225 & 0.0435\\
MSE                 & 0.0494  & 0.0493  & 0.0508  & 0.0872 && 0.0356 & 0.0355 & 0.0356 & 0.0660  && 0.0225 & 0.0225 & 0.0225 & 0.0435\\\\
$\widehat{\beta_2}$ & 1.0040 & 1.0047 & 1.0044 & 1.0017 && 1.0027 & 1.0017 & 1.0013 & 1.0006 && 0.9999 & 0.9967 & 0.9963 & 0.9971\\
Bias                & 0.0040 & 0.0047 & 0.0044 & 0.0017 && 0.0027 & 0.0017 & 0.0013 & 0.0006 && $-$0.0001 & $-$0.0033 & $-$0.0037 & $-$0.0029\\
variance            & 0.0410 & 0.0409 & 0.0420 & 0.0721 && 0.0340 & 0.0339 & 0.0341 & 0.0618 && 0.0290 & 0.0290 & 0.0293 & 0.0549\\
MSE                 & 0.0410 & 0.0409 & 0.0420 & 0.0721 && 0.0340 & 0.0339 & 0.0341 & 0.0618 && 0.0290 & 0.0290 & 0.0293 & 0.0549\\\\
$\widehat{\nu_0}$  & 1.1367 & 1.0232 & 1.0213 & 1.0123 && 1.0753 & 1.0068 & 1.0112 & 0.9864  && 1.0602 & 1.0088 & 1.0100 & 0.9980 \\
Bias               & 0.1367 & 0.0232 & 0.0213 & 0.0123 && 0.0753 & 0.0068 & 0.0112 & $-$0.0136 && 0.0602 & 0.0088 & 0.0100 & $-$0.0020\\
variance           & 0.6454 & 0.5492 & 0.7384 & 1.1125 && 0.6011 & 0.5218 & 0.7964 & 1.0588  && 0.3988 & 0.3587 & 0.4839 & 0.7352\\
MSE                & 0.6641 & 0.5498 & 0.7388 & 1.1127 && 0.6068 & 0.5219 & 0.7966 & 1.0590  && 0.4025 & 0.3588 & 0.4840 & 0.7352\\\\
$\widehat{\nu_1}$  & 1.1884 & 1.0039 & 0.9675  & 1.0051 && 1.1069 & 1.0147 & 0.9922 & 1.0119  && 1.0253 & 0.9927 & 0.9881 & 1.0021\\
Bias               & 0.1884 & 0.0039 & $-$0.0325 & 0.0051 && 0.1069 & 0.0147 & $-$0.0078 & 0.0119 && 0.0253 & $-$0.0073 & $-$0.0119 & 0.0021\\
variance           & 1.5315 & 1.2260 & 2.1793  & 2.5333 && 0.8943 & 0.7649 & 0.9137 & 1.5950  && 0.5878 & 0.5286 & 0.6083 & 1.0811\\
MSE                & 1.5670 & 1.2260 & 2.1804  & 2.5333 && 0.9058 & 0.7652 & 0.9137 & 1.5952  && 0.5885 & 0.5286 & 0.6085 & 1.0811\\\\
$\widehat{\nu_2}$  & 1.0311 & 1.0220 & 1.0017 & 0.9895 && 1.0974 & 1.0113 & 0.9926 & 1.0143  && 1.0901 & 1.0082 & 0.9948 & 1.0039\\
Bias               & 0.0311 & 0.0220 & 0.0017 & $-$0.0105&& 0.0974 & 0.0113 & $-$0.0074 & 0.0143 && 0.0901 & 0.0082 & $-$0.0052 & 0.0039\\
variance           & 1.3491 & 1.0920 & 1.5705 & 2.2549 && 1.0802 & 0.9325 & 1.3096 & 1.8828  && 0.7401 & 0.6612 & 0.8356 & 1.3574\\
MSE                & 1.3500 & 1.0925 & 1.5705 & 2.2550 && 1.0897 & 0.9326 & 1.3097 & 1.8830  && 0.7482 & 0.6613 & 0.8356 & 1.3574\\
\hline
\end{tabular}
}
\label{tab2}
\end{table}

%%%%%%%%%%%%%%%%%%%%%%%%%%%%%%%%%%%%%%%%%%%%%%%%%%%%%%%%%%%%%%%%%%%%%%%%%%%%%%
%              resultados p=q=3
%%%%%%%%%%%%%%%%%%%%%%%%%%%%%%%%%%%%%%%%%%%%%%%%%%%%%%%%%%%%%%%%%%%%%%%%%%%%%%

\begin{table}[!htb]
  \centering
\caption{Simulation results for $p=q=3$.}
\resizebox{\linewidth}{!}{
\begin{tabular}{lrrrrrrrrrrrrrr}
\hline
 & \multicolumn{4}{c}{$n=30$} && \multicolumn{4}{c}{$n=40$} && \multicolumn{4}{c}{$n=60$}\\
  \cline{2-5} \cline{7-10} \cline{12-15}
Estimates & MLE &Cox-Snell &Firth &p-boot && MLE &Cox-Snell & Firth & p-boot && MLE & Cox-Snell &Firth &p-boot \\ \hline
$\widehat{\beta_0}$ & 0.9959 & 0.9989 & 1.0028 & 0.9973    && 0.9985 & 1.0002 & 1.0017 & 1.0011  && 0.9973 & 1.0011 & 1.0023 & 0.9994\\
Bias                & $-$0.0041 & $-$0.0011 & 0.0028 & $-$0.0027 && $-$0.0015 & 0.0002 & 0.0017 & 0.0011 && $-$0.0027 & 0.0011 & 0.0023 & $-$0.0006\\
variance            & 0.0249 & 0.0249 & 0.0267 & 0.0418    && 0.0301 & 0.0301 & 0.0324 & 0.0531  && 0.0111 & 0.0111 & 0.0112 & 0.0204 \\
MSE                 & 0.0249 & 0.0249 & 0.0267 & 0.0418    && 0.0301 & 0.0301 & 0.0324 & 0.0531  && 0.0111 & 0.0111 & 0.0112 & 0.0204\\\\
$\widehat{\beta_1}$ & 1.0026 & 1.0024 & 1.0016 & 1.0031 && 0.9983 & 0.9996 & 1.0002 & 0.9975    && 1.0018 & 1.0016 & 1.0015 & 1.0014\\
Bias                & 0.0026 & 0.0024 & 0.0016 & 0.0031 && $-$0.0017 & $-$0.0004 & 0.0002 & $-$0.0025 && 0.0018 & 0.0016 & 0.0015 & 0.0014\\
variance            & 0.0354 & 0.0354 & 0.0372 & 0.0598 && 0.0249 & 0.0249 & 0.0261 & 0.0443    && 0.0135 & 0.0135 & 0.0138 & 0.0249\\
MSE                 & 0.0354 & 0.0354 & 0.0372 & 0.0598 && 0.0249 & 0.0249 & 0.0261 & 0.0443    && 0.0136 & 0.0135 & 0.0138 & 0.0249\\\\
$\widehat{\beta_2}$ & 1.0019 & 1.0027 & 1.0018 & 1.0047 && 0.9995 & 0.9998 & 0.9995 & 0.9993     && 0.9993 & 0.9975 & 0.9972 & 0.9994\\
Bias                & 0.0019 & 0.0027 & 0.0018 & 0.0047 && $-$0.0005 & $-$0.0002 & $-$0.0005 & $-$0.0007 && $-$0.0007 & $-$0.0025 & $-$0.0028 & $-$0.0006\\
variance            & 0.0309 & 0.0309 & 0.0327 & 0.0528 && 0.0235 & 0.0234 & 0.0245 & 0.0413     && 0.0167 & 0.0167 & 0.0171 & 0.0307\\
MSE                 & 0.0309 & 0.0309 & 0.0327 & 0.0528 && 0.0235 & 0.0234 & 0.0245 & 0.0413     && 0.0167 & 0.0167 & 0.0171 & 0.0307\\\\
$\widehat{\beta_3}$ & 0.9937 & 0.9943 & 0.9942 & 0.9947     && 0.9978 & 0.9987 & 0.9983 & 0.9997     && 0.9997 & 0.9980 & 0.9974 & 0.9990\\
Bias                & $-$0.0063 & $-$0.0057 & $-$0.0058 & $-$0.0053 && $-$0.0022 & $-$0.0013 & $-$0.0017 & $-$0.0003 && $-$0.0003 & $-$0.0020 & $-$0.0026 & $-$0.0010\\
variance            & 0.0397 & 0.0397 & 0.0419 & 0.0659     && 0.0309 & 0.0309 & 0.0336 & 0.0556     && 0.0190 & 0.0190 & 0.0192 & 0.0347\\
MSE                 & 0.0398 & 0.0397 & 0.0420 & 0.0659     && 0.0309 & 0.0309 & 0.0336 & 0.0556     && 0.0190 & 0.0190 & 0.0192 & 0.0347\\\\
$\widehat{\nu_0}$  & 1.1515 & 1.0091 & 1.0325 & 1.0023 && 1.0786 & 1.0051 & 1.0316 & 0.9976  && 1.0479 & 1.0049 & 1.0138 & 0.9899\\
Bias               & 0.1515 & 0.0091 & 0.0325 & 0.0023 && 0.0786 & 0.0051 & 0.0316 & $-$0.0024 && 0.0479 & 0.0049 & 0.0138 & $-$0.0101\\
variance           & 0.5818 & 0.4711 & 1.6123 & 0.9480 && 0.4827 & 0.4123 & 0.8403 & 0.8484  && 0.3339 & 0.3001 & 0.6093 & 0.5945\\
MSE                & 0.6048 & 0.4712 & 1.6134 & 0.9480 && 0.4889 & 0.4123 & 0.8413 & 0.8484  && 0.3362 & 0.3001 & 0.6095 & 0.5946\\\\
$\widehat{\nu_1}$  & 1.1387 & 0.9982 & 0.9634 & 0.9994    && 1.1418 & 0.9729 & 0.9457 & 0.9954    && 1.0161 & 0.9977 & 0.9981 & 1.0051 \\
Bias               & 0.1387 & $-$0.0018 & $-$0.0366 & $-$0.0006 && 0.1418 & $-$0.0271 & $-$0.0543 & $-$0.0046 && 0.0161 & $-$0.0023 & $-$0.0019 & 0.0051 \\
variance           & 1.2110 & 0.9163 & 2.4586 & 1.8249    && 0.7643 & 0.6403 & 1.3250 & 1.3258    && 0.4878 & 0.4270 & 0.5802 & 0.8715\\
MSE                & 1.2303 & 0.9164 & 2.4599 & 1.8249    && 0.7844 & 0.6411 & 1.3280 & 1.3259    && 0.4881 & 0.4270 & 0.5802 & 0.8715\\\\
$\widehat{\nu_2}$  & 1.0566 & 1.0589 & 1.0382 & 1.0115 && 1.0808 & 0.9918 & 0.9678 & 1.0014   && 1.0969 & 1.0066 & 0.9867 & 1.0172\\
Bias               & 0.0566 & 0.0589 & 0.0382 & 0.0115 && 0.0808 & $-$0.0082 & $-$0.0322 & 0.0014 && 0.0969 & 0.0066 & $-$0.0133 & 0.0172 \\
variance           & 1.1501 & 0.8767 & 1.5965 & 1.7398 && 0.8195 & 0.6709 & 1.3184 & 1.3820   && 0.5622 & 0.4854 & 0.8178 & 0.9851\\
MSE                & 1.1533 & 0.8802 & 1.5979 & 1.7400 && 0.8260 & 0.6710 & 1.3195 & 1.3820   && 0.5716 & 0.4855 & 0.8180 & 0.9854\\\\
$\widehat{\nu_3}$  & 1.1683 & 1.0215 & 0.9504 & 1.0233  && 1.0929 & 1.0866 & 1.0413 & 1.0185  && 1.0920 & 1.0114 & 0.9911 & 0.9990\\
Bias               & 0.1683 & 0.0215 & $-$0.0496 & 0.0233 && 0.0929 & 0.0866 & 0.0413 & 0.0185  && 0.0920 & 0.0114 & $-$0.0089 & $-$0.0010\\
variance           & 1.0738 & 0.8799 & 2.3169 & 1.7278  && 0.9323 & 0.7420 & 2.0245 & 1.4912  && 0.4791 & 0.4213 & 0.5501 & 0.8672\\
MSE                & 1.1021 & 0.8803 & 2.3194 & 1.7283  && 0.9409 & 0.7495 & 2.0262 & 1.4916  && 0.4876 & 0.4214 & 0.5502 & 0.8672\\
\hline
\end{tabular}
}
\label{tab3}
\end{table}

\begin{figure}
\subfigure{
\includegraphics[scale=0.245]{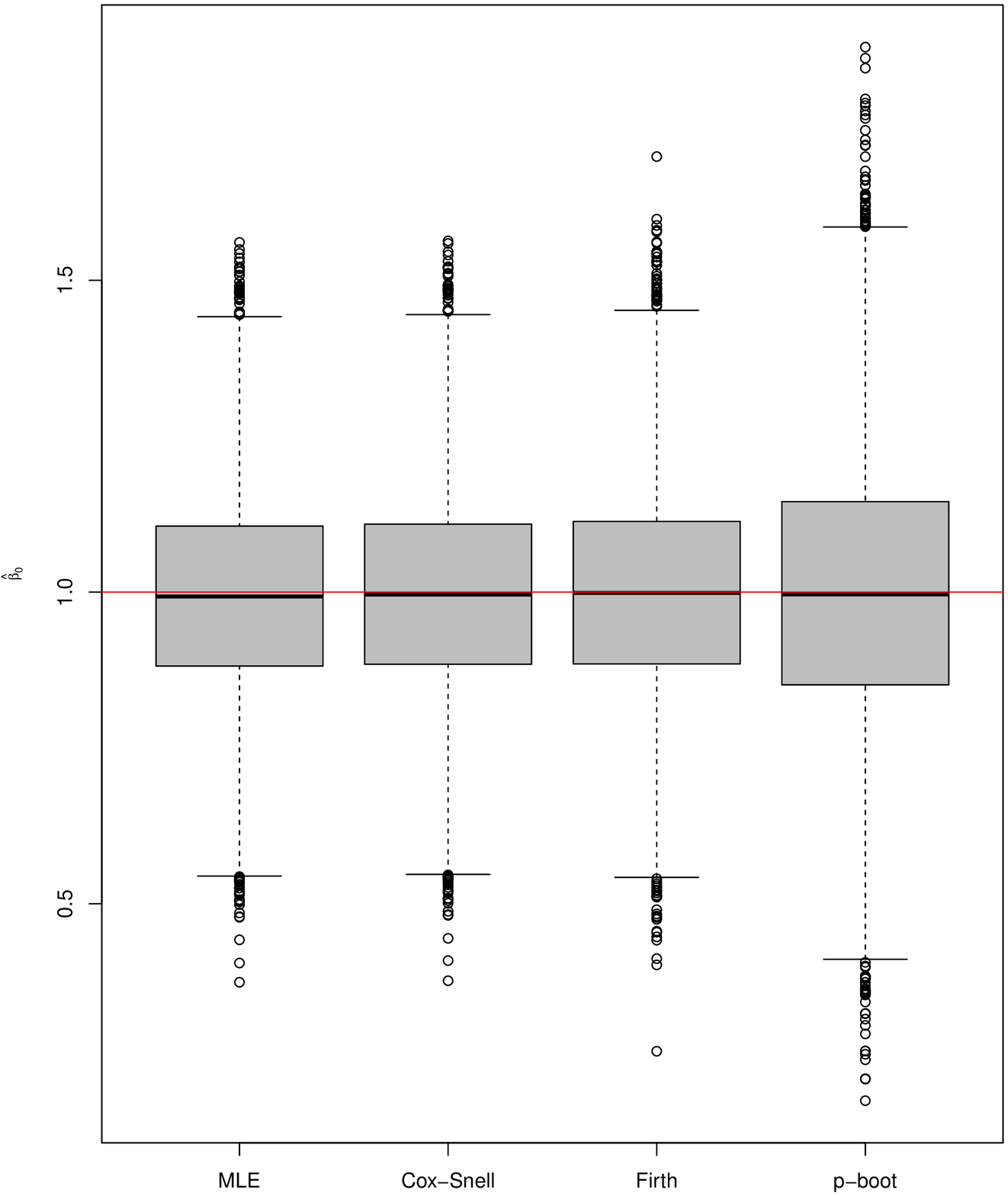}}
\subfigure{
\includegraphics[scale=0.245]{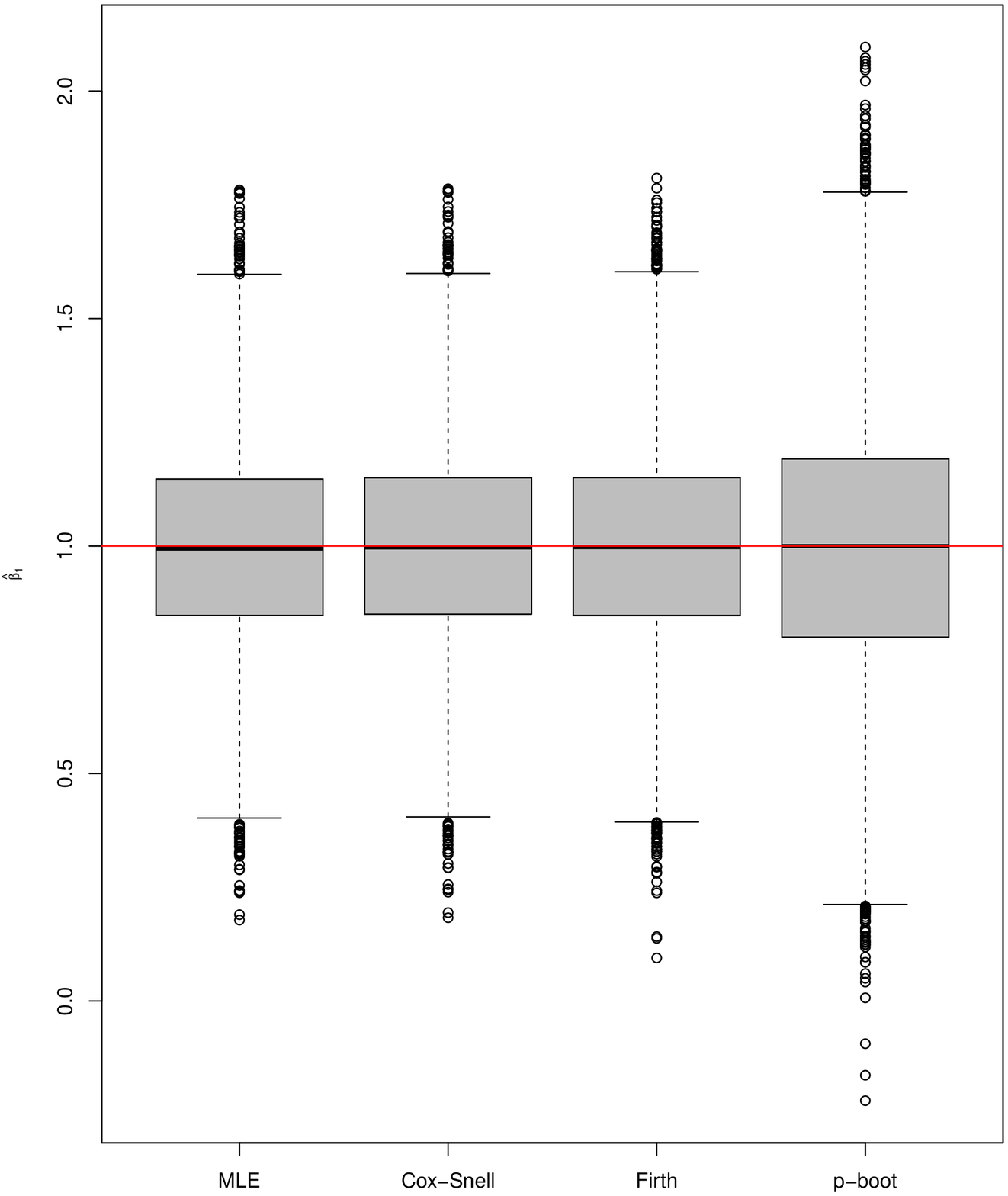}}
\subfigure{
\includegraphics[scale=0.245]{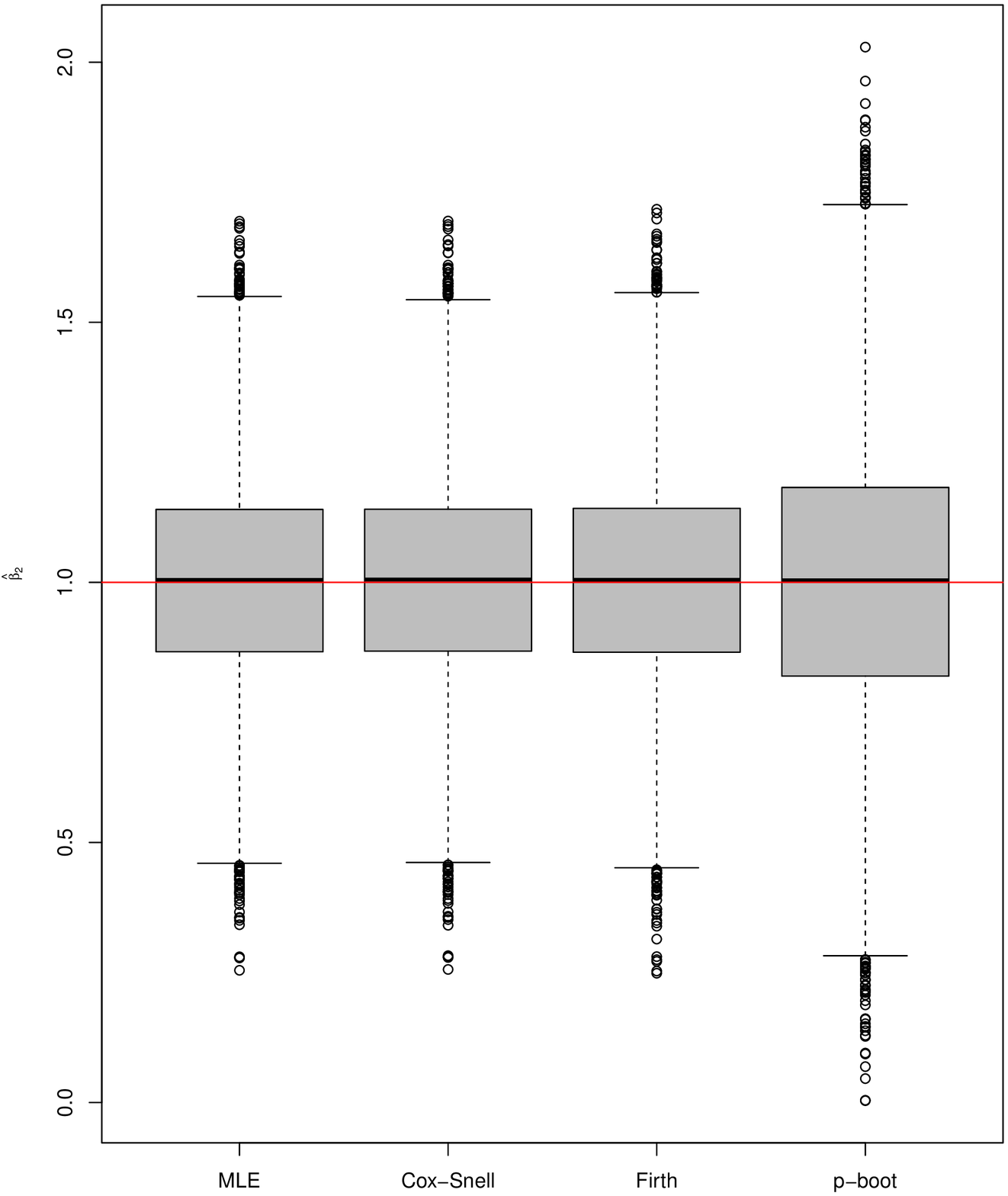}}
%%%%%%%%%%%%%%%%%%%%%%%%%%%%%%%%%%%%%%%%%%%%%%%%%%%%%%
\subfigure{
\includegraphics[scale=0.245]{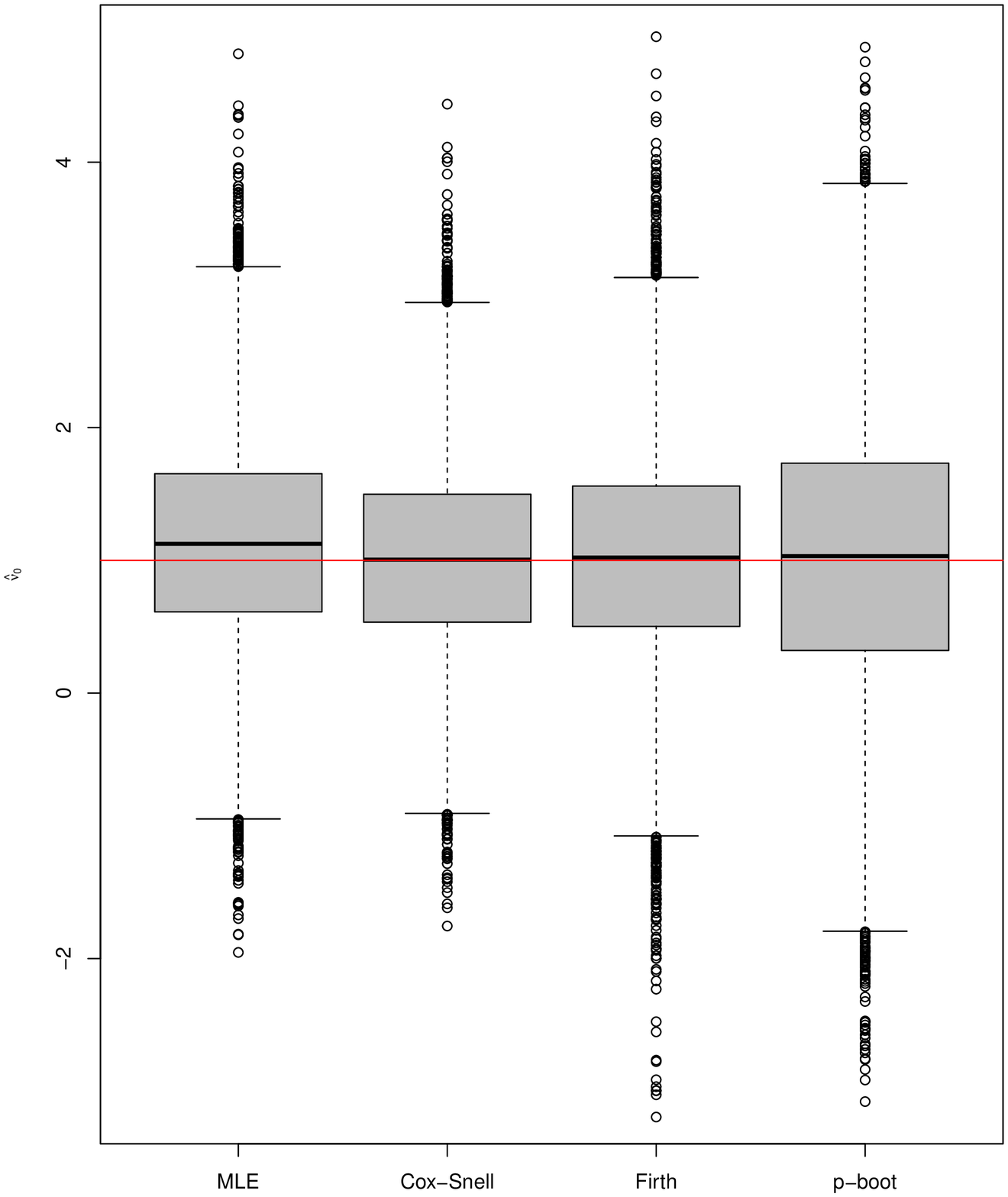}}
\subfigure{
\includegraphics[scale=0.245]{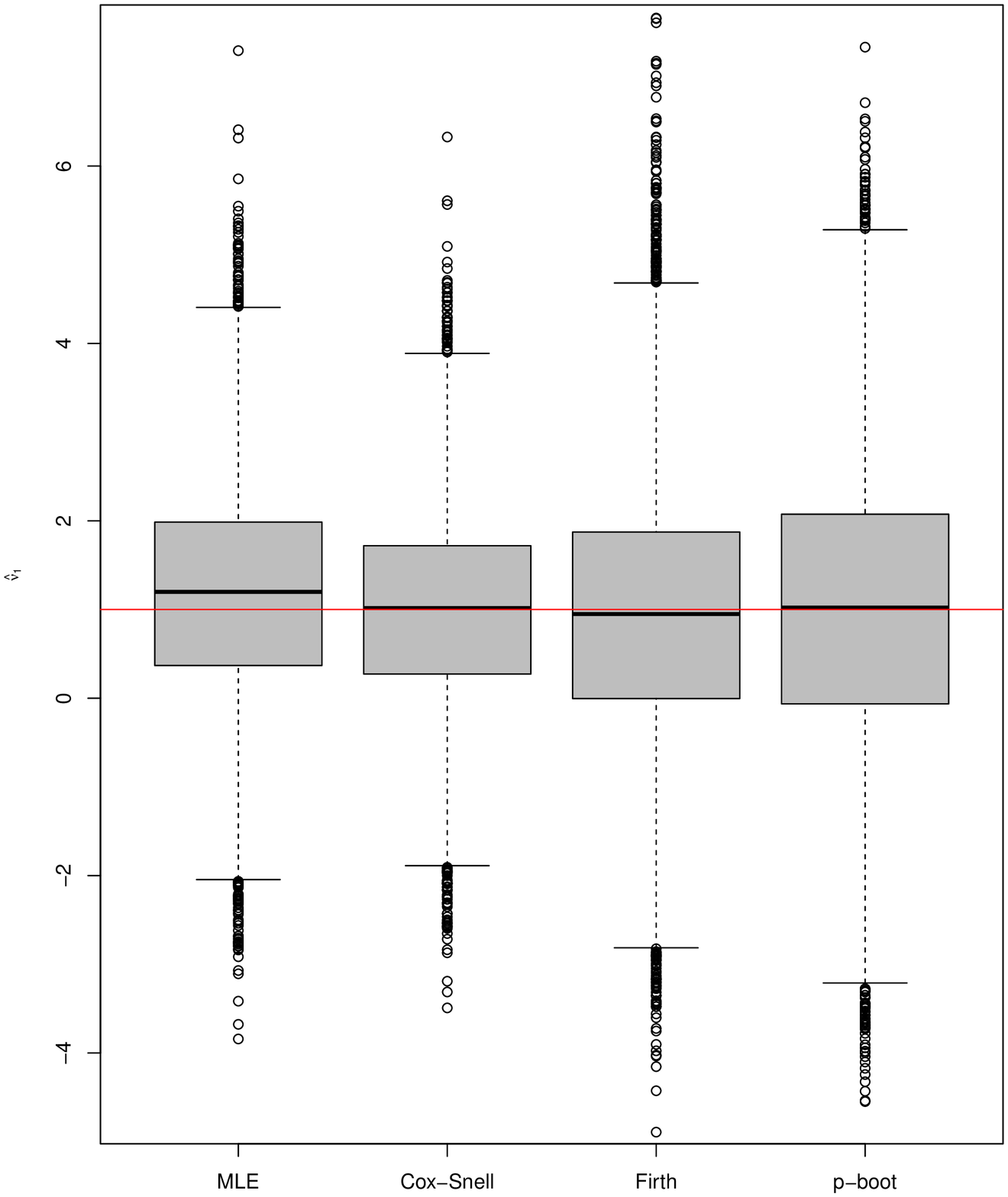}}
\subfigure{
\includegraphics[scale=0.245]{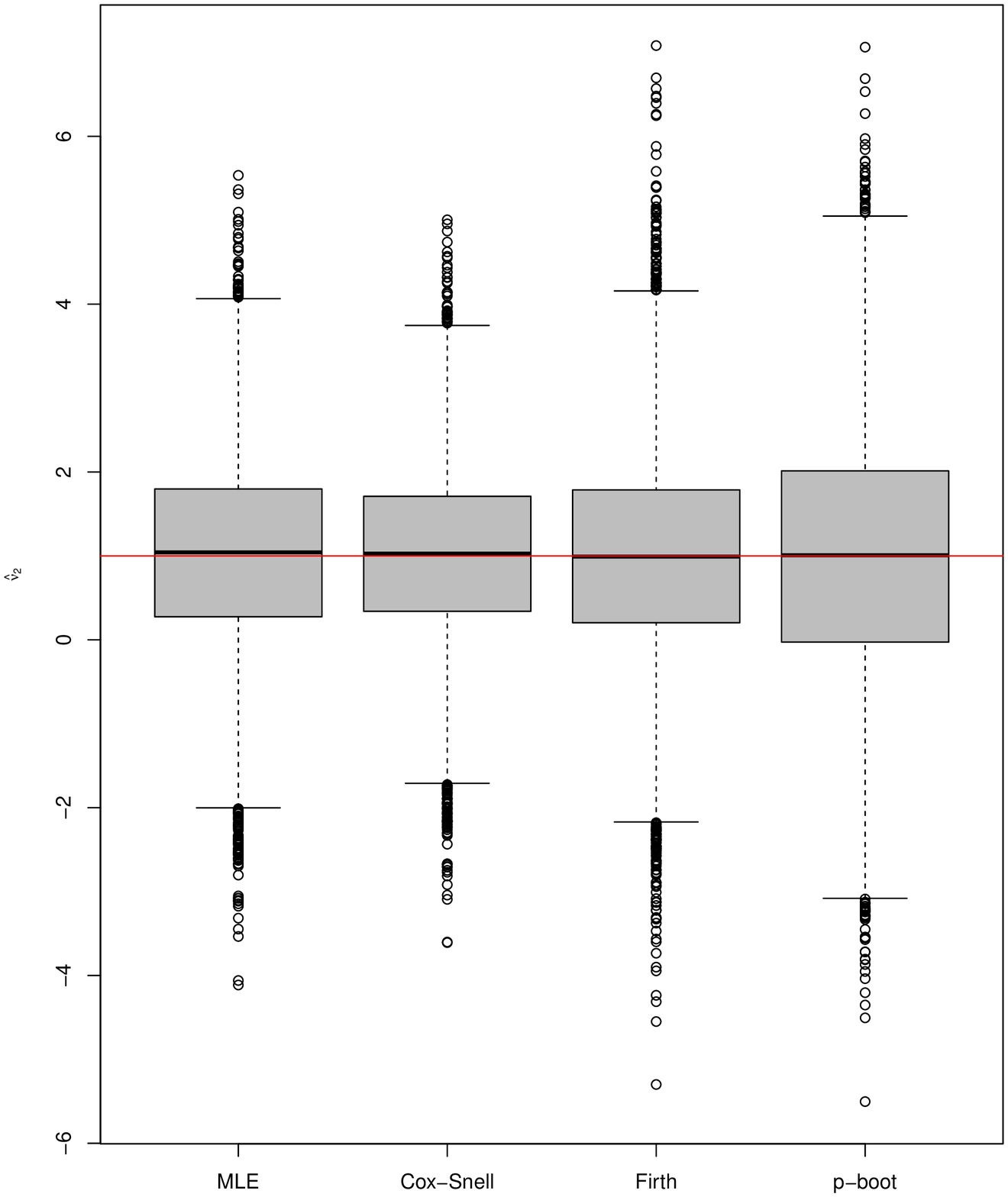}}
%%%%%%%%%%%%%%%%%%%%%%%%%%%%%%%%%%%%%%%%%%%%%%%%%%%%%%%%%%
\caption{Box plots from 10.000 simulated estimates of $\beta_0$, $\beta_1$, $\beta_2$, $\nu_0$, $\nu_1$ and $\nu_2$ for $n = 30$.}\label{fig:boxplot}
\end{figure}

\section{An application}\label{sec5}
\noindent

In order to illustrate the proposed methodology, in this section, we apply the estimation methods considered in the previous section to a real situation. We consider the real dataset used in \cite{Bonnail}. The main purpose is to assess sediment quality using the freshwater clam Corbiculafluminea to determine its adequacy as a biomonitoring tool in relation to theoretical risk indexes and regulatory thresholds. The study contains 27 observations (small-sized sample), which measured, among other characteristics, the dry weight tissue of the clams (\texttt{dry}, in g), wet weight tissue (\texttt{wet}, in g), and the concentrations of caesium (\texttt{cs}) in the soft tissue. Such minerals were considered in 100 micrograms per liter ($100\mu g /L$).

We adopted a BP model to fit the dry weight tissue of the clams; that is,
we consider that \texttt{dry}$_i \sim$ BP$(\mu_i,\phi_i)$ with systematic components given by
\begin{align*}
\log(\mu_i) & = \beta_0 + \beta_1\texttt{wet}_i + \beta_2\texttt{cs}_i \nonumber \\
\log(\phi_i)& = \nu_0 + \nu_1\texttt{wet}_i, \quad i=1, \ldots, 27. \label{cov.app1}
\end{align*}

An R implementation for obtaining MLEs along with their corrected versions proposed in this article related to the data used is available at GitHub BiasBPR\footnote{\url{https://github.com/sesiommedeiros/BiasBPR}} repository.

Table \ref{tabaplic1} presents the maximum likelihood estimates along with their corrected versions and the corresponding estimates of asymptotic standard errors in parentheses. Note that for the parameters that model the precision, $\widehat{\nu}_{0}$ and $\widehat{\nu}_{1},$ the maximum likelihood estimates are smaller than the bias corrected ones, while for the parameters that model the mean, $\widehat{\beta}_{0},$ $\widehat{\beta}_{1}$ and $\widehat{\beta}_{2},$ the estimates are quite close.

%The adjusted versions are the corrective bias-corrected estimator (Cox-Snell), the preventive bias-corrected estimator (Firth), and the parametric bootstrap %bias-corrective estimator (p-boot).

\begin{table}[!htb]
\centering
\caption{Estimated values of the parameters with estimated asymptotic standard errors in parenthesis.}
\begin{tabular}{lrrrr}
  \hline
 Estimates              & MLE & Cox-Snell & Firth & p-boot \\
  \hline
$\widehat{\beta}_{0}$&$-$1.5550 &  $-$1.5550& $-$1.5596& $-$1.5526 \\
 &(0.0224) &   (0.0252) &(0.0257)& (0.0254)\\
$\widehat{\beta}_{1}$&$-$0.0221  & $-$0.0221& $-$0.0193& $-$0.0231\\
 &(0.0105)  &  (0.0118) &(0.0121)& (0.0120)\\
$\widehat{\beta}_{2}$&$-$0.0182 &  $-$0.0183 &$-$0.0287& $-$0.0203\\
 &(0.1243)  &  (0.1393) &(0.1418) &(0.1406)\\
$\widehat{\nu}_{0}$ &1.4536 &   1.5362& 1.5881& 1.5950\\
 &(0.9059)  &  (0.9060) &(0.9060) &(0.9060)\\
$\widehat{\nu}_{1}$& 5.1014 &   4.9186& 4.8675& 4.8728\\
 &(0.5896) &   (0.5897) & (0.5897)& (0.5897)\\ \hline
\end{tabular}
\label{tabaplic1}
\end{table}

In the Table \ref{tabaplic2}, we present the relative changes (RCs). The RCs are
calculated from $\textrm{RC}(\widehat\theta) = |(\widehat\theta - \widehat{\theta}_{\textrm{o}})/\widehat{\theta}_{\textrm{o}}| \times 100\%$,
where $\widehat\theta$ denotes the MLE of $\theta$ and $\widehat{\theta}_{\textrm{o}}$ denotes the bias-corrected MLE of $\theta$.
From Table \ref{tabaplic2}, the bias-corrected MLEs for $\nu_0$ and $\nu_1$ present similar results.
In contrast to the Cox-Snell and p-boot bias-corrected estimators, the preventive method (Firth) gives estimates that dramatically change for $\beta_1$ and $\beta_2$. For example, the second-order bias is 36.585\% of the total amount of the MLE of $\beta_2$. Thus, this real example illustrates that bias corrections can have a great effect on the conclusions.

\begin{table}[!htb]
\centering
\caption{Relative changes for each parameter.}
\begin{tabular}{lccccc}
  \hline
Estimator&    RC($\widehat{\beta}_{0}$)&   RC($\widehat{\beta}_{1}$)&  RC($\widehat{\beta}_{2}$)&  RC($\widehat{\nu}_{0}$)& RC($\widehat{\nu}_{1}$)\\ \hline
Cox-Snell& 0.0000  & 0.0000  & 0.5464&  5.3769&  3.7165\\
Firth    & 0.2950  & 14.508  & 36.585&  8.4692&  4.8053\\
p-boot   & 0.1546  & 4.3290  & 10.345&  8.8652&  4.6914\\
\hline
\end{tabular}
\label{tabaplic2}
\end{table}

\section{Concluding remarks}\label{sec6}
\noindent

In this paper, we have examined a wide range of estimators for the unknown parameter vector of the BP regression model.
In particular,  we have derived a closed-form expression, in matrix form, for the second order biases of the ML estimators of the parameters that index the BP regression model proposed by \cite{bour18}. For this, we use the expressions obtained through Cox and Snell's \citep{coxsnell68} formulae and Firth's \citep{Firth1993} estimating equation. We also considered a bias correction based on parametric bootstrap. The numerical evidence here presented shows that our proposed estimators has good finite-sample behavior, even when the sample size is small. For the mean structure, we observe that the MLE presents a very small bias  (even in small samples). In this case, it is not necessary to use the bias-corrected estimators. However, for the precision structure, we observe
that the MLE can become considerably biased and, therefore, we strongly recommend its bias correction. This behavior was also observed in the application to the real data set presented, therefore, we strongly recommend that practitioners use these corrected estimators when modeling data using the BP regression model. Finally, we have applied our proposed estimators to a real data.


%\bibliographystyle{authordate1}
%\bibliography{reference}


\section*{Appendix A}\label{secA}
\noindent

In this Appendix, we presente the Fisher's information matrix for the BP regression model, which is expressed in matrix form as
\begin{equation}\label{eqm22}
\mathbf{K}(\bm{\beta}, \bm{\nu})=\left(
\begin{array}{cc}
  \mathbf{X}^{\top} K_{\beta\beta} \mathbf{X} &   \mathbf{X}^{\top} K_{\beta\nu} \mathbf{Z} \\
 \mathbf{Z}^{\top} K_{\nu\beta} \mathbf{X} &  \mathbf{Z}^{\top} K_{\nu\nu} \mathbf{Z}  \\
\end{array}
\right),
\end{equation}
where
\begin{eqnarray*}
K_{\beta\beta} &=& \diag\left\lbrace (1 + \phi_i)^2 a_i \left(\frac{\p \mu_i}{\p \eta_{1i}}\right)^2\right\rbrace, \hfill
K_{\nu\nu} = \diag \left\lbrace b_i \left(\frac{\p \phi_i}{\p \eta_{2i}} \right)^2\right\rbrace,\\
K_{\beta\nu} &=& K_{\nu\beta}^\top = \diag\left\lbrace (1 + \phi_i) [a_i \mu_i - \psi^{(1)}(\mu_i(1 + \phi_i) + \phi_i + 2)] \frac{\p \mu_i}{\p \eta_{1i}}\frac{\p \phi_i}{\p \eta_{2i}}\right\rbrace ,\\
\end{eqnarray*}
with $a_i = \psi^{(1)}(\mu_i(1+\phi_i)) - \psi^{(1)}(\mu_i(1+\phi_i) + \phi_i + 2)$ and $b_i = \mu_i^2 \psi^{(1)}(\mu_i(1+\phi_i)) - (1 + \mu_i)^2 \psi^{(1)}(\mu_i(1+\phi_i) + \phi_i + 2) + \psi^{(1)}(\phi_1 + 2)$.

We can rewrite the Fisher's information matrix given in (\ref{eqm22}). For this, let $\widetilde{\bm{K}}$ be a $2n \times 2n$ matrix and $\mathbb{X}$ be a $2n \times (p+q)$ matrix defined, respectively, as
\begin{eqnarray*}
\widetilde{\bm{K}} &=& \left(
\begin{array}{cc}
     K_{\beta\beta}& K_{\beta\nu} \\
     K_{\nu\beta}  & K_{\nu\nu}
\end{array}
\right).
\end{eqnarray*}
and
\begin{eqnarray*}
\mathbb{X} &=& \left(
\begin{array}{cc}
     \bm{X}& \bm{0} \\
     \bm{0}& \bm{Z}
\end{array}
\right).
\end{eqnarray*}
Thus, we have that $$\mathbf{K}(\bm{\beta}, \bm{\nu})= \mathbb{X}^\top \widetilde{K}\mathbb{X}.$$

\section*{Appendix B}\label{secB}
\noindent

In this Appendix, we present the cumulants and derivatives needed to obtain the second order bias of $\bm{\widehat{\beta}}$ and $\bm{\widehat{\nu}}.$ In addiction, we describe in details how to obtain $B(\bm{\widehat{\beta}})$ and $B(\bm{\widehat{\nu}})$ from Eq. (\ref{vcs}). In order to present the cumulants in a summarized form, consider the following quantities:

\begin{eqnarray*}
%\mu_i^* &=& \psi^{(0)}(\mu_i(1+\phi_i)) - \psi^{(0)}(\mu_i(1+\phi_i) + \phi_i + 2), \\
		a_i &=& \psi^{(1)}(\mu_i(1+\phi_i)) - \psi^{(1)}(\mu_i(1+\phi_i) + \phi_i + 2), \\
		b_i &=& \mu_i^2 \psi^{(1)}(\mu_i(1+\phi_i)) - (1 + \mu_i)^2 \psi^{(1)}(\mu_i(1+\phi_i) + \phi_i + 2) + \psi^{(1)}(\phi_1 + 2),\\
		c_i &=& \psi^{(2)}(\mu_i(1+\phi_i)) - \psi^{(2)}(\mu_i(1+\phi_i) + \phi_i + 2), \\
		d_i &=& (1 + \mu_i)^2 \psi^{(2)}(\mu_i(1+\phi_i) + \phi_i + 2) - \mu_i^2 \psi^{(2)}(\mu_i(1+\phi_i)), \\
		e_i &=& (1 + \mu_i)^3 \psi^{(2)}(\mu_i(1+\phi_i) + \phi_i + 2) - \mu_i^3 \psi^{(2)}(\mu_i(1+\phi_i)) - \psi^{(2)}(\phi_i + 2).
\end{eqnarray*}
So, the cumulants needed here are presented as follow:
\begin{eqnarray*}
\kappa_{rs} &=& - \sum_{i=1}^n (1 + \phi_i)^2 a_i \left(\frac{\p \mu_i}{\p \eta_{1i}}\right)^2 x_{ir}x_{is},\\
\kappa_{rS} &=& - \sum_{i=1}^n (1 + \phi_i) [a_i \mu_i - \psi^{(1)}(\mu_i(1 + \phi_i) + \phi_i + 2)] \frac{\p \mu_i}{\p \eta_{1i}}\frac{\p \phi_i}{\p \eta_{2i}} x_{ir}z_{iS},\\
\kappa_{RS} &=& - \sum_{i=1}^n b_i \left(\frac{\p \phi_i}{\p \eta_{2i}} \right)^2 z_{iR}z_{iS},\\
\kappa_{rsu} &=& -\sum_{i=1}^n (1 + \phi_i)^2 \left \{(1 + \phi_i)c_i \left(\frac{\p \mu_i}{\p \eta_{1i}} \right)^3 + 3a_i \frac{\p \mu_i}{\p \eta_{1i}} \frac{\p^2 \mu_i}{\p \eta^2_{1i}} \right \} x_{ir}x_{is}x_{iu},\\
\kappa_{rsU} &=& - \sum_{i=1}^n (1 + \phi_i) \left\{2a_i + (1 + \phi_i)c_i \mu_i - (1 + \phi_i)\psi^{(2)}(\mu_i(1+\phi_i) + \phi_i + 2) \right\} \left(\frac{\p \mu_i}{\p \eta_{1i}} \right)^2 \frac{\p \phi_i}{\p \eta_{2i}} x_{ir}x_{is}z_{iU}\\
				&+& \sum_{i=1}^n (1 + \phi_i) \left\{\psi^{(1)}(\mu_i(1+\phi_i) + \phi_i + 2) - a_i\mu_i \right\}\frac{\p^2 \mu_i}{\p \eta^2_{1i}}\frac{\p \phi_i}{\p \eta_{2i}}x_{ir}x_{is}z_{iU},\\
\kappa_{rSU} &=& \sum_{i=1}^n \{\psi^{(1)}(\mu_i(1+\phi_i) + \phi_i + 2) - a_i\mu_i \} \left[(1 + \phi_i)\frac{\p \mu_i}{\p \eta_{1i}}\frac{\p^2 \phi_i}{\p \eta^2_{2i}} + 2 \left( \frac{\p \phi_i}{\p \eta_{2i}} \right)^2 \frac{\p \mu_i}{\p \eta_{1i}} \right] x_{ir}z_{iS}z_{iU}\\
				&+& \sum_{i=1}^n (1 + \phi_i) d_i \frac{\p \mu_i}{\p \eta_{1i}} \left( \frac{\p \phi_i}{\p \eta_{2i}} \right)^2 x_{ir}z_{iS}z_{iU},\\
\kappa_{RSU} &=& \sum_{i=1}^n \left\{e_i \left(\frac{\p \phi_i}{\p \eta_{2i}} \right)^3 - 3b_i \frac{\p^2 \phi_i}{\p \eta^2_{2i}}\frac{\p \phi_i}{\p \eta_{2i}} \right\} z_{iR}z_{iS}z_{iU}.
\end{eqnarray*}

Taking the derivative of the cumulants with respect to the model parameters, we have
\begin{eqnarray*}
\kappa_{rs}^{(u)} &=& - \sum_{i=1}^n (1 + \phi_i)^2 \left \{(1 +  \phi_i)c_i \left(\frac{\p \mu_i}{\p \eta_{1i}}\right)^3 + 2a_i \frac{\p \mu_i}{\p \eta_{1i}} \frac{\p^2 \mu_i}{\p \eta^2_{1i}}  \right \}x_{ir}x_{is}x_{iu},\\
\kappa_{rs}^{(U)} &=& - \sum_{i=1}^n \left\{ (1 + \phi_i)^2[c_i \mu_i - \psi^{(2)}(\mu_i(1+\phi_i) + \phi_i + 2)] + 2(1 + \phi_i)a_i \right\} \left(\frac{\p \mu_i}{\p \eta_{1i}} \right)^2 \frac{\p \phi_i}{\p \eta_{2i}}x_{ir}x_{is}z_{iU},\\
\kappa_{RS}^{(u)} &=& \sum_{i=1}^n \left\{d_i(1 + \phi_i) + 2 \psi^{(1)}(\mu_i(1+\phi_i) + \phi_i + 2) - 2a_i \mu_i \right\}\frac{\p \mu_i}{\p \eta_{1i}}\left(\frac{\p \phi_i}{\p \eta_{2i}}\right)^2 z_{iR}z_{iS}x_{iu},\\
\kappa_{RS}^{(U)} &=& \sum_{i=1}^n \left\{e_i\left( \frac{\p \phi_i}{\p \eta_{2i}} \right)^3 - 2b_i \frac{\p \phi_i}{\p \eta_{2i}}\frac{\p^2 \phi_i}{\p \eta^2_{2i}} \right\}z_{iR}z_{iS}z_{iU},\\
\kappa_{rS}^{(u)} &=& - \sum_{i=1}^n (1+\phi_i) \left\{(1+\phi_i)\mu_i c_i + a_i - (1+\phi_i)\psi^{(2)}(\mu_i(1+\phi_i) + \phi_i + 2)  \right\} \left(\frac{\p \mu_i}{\p \eta_{1i}} \right)^2 \frac{\p \phi_i}{\p \eta_{2i}} x_{ir}x_{iu}z_{iS}\\
						&-& \sum_{i=1}^n (1 + \phi_i) \left\{a_i \mu_i - \psi^{(1)}(\mu_i(1+\phi_i) + \phi_i + 2) \right\} \frac{\p^2 \mu_i}{\p \eta_{1i}^2} \frac{\p \phi_i}{\p \eta_{2i}}x_{ir}x_{iu}z_{iS},\\
\kappa_{rS}^{(U)} &=& \sum_{i=1}^n \left\{(1+\phi_i)d_i - a_i \mu_i + \psi^{(1)}(\mu_i(1+\phi_i) + \phi_i + 2) \right \} \frac{\p \mu_i}{\p \eta_{1i}} \left(\frac{\p \phi_i}{\p \eta_{2i}} \right)^2 x_{ir}z_{iS}z_{iU}\\
						&+& \sum_{i=1}^n (1 + \phi_i) \left\{\psi^{(1)}(\mu_i(1+\phi_i) + \phi_i + 2) - a_i\mu_i  \right\} \frac{\p \mu_i}{\p \eta_{1i}} \frac{\p^2 \phi_i}{\p \eta^2_{2i}} x_{ir}z_{iS}z_{iU}.
\end{eqnarray*}
Consider now the following diagonal matrices:
\begin{eqnarray*}
M_1 &=& \diag \left(- \frac{(1+\phi_i)^2}{2} \left[ (1+\phi_i)c_i \left(\frac{\p \mu_i}{\p \eta_{1i}} \right)^3 + a_i\frac{\p \mu_i}{\p \eta_{1i}}\frac{\p^2 \mu_i}{\p \eta_{1i}^2} \right] \right), \\
M_2 &=& \diag \left( \frac{(1+\phi_i)}{2} \left[-a_i \mu_i + \psi^{(1)}(\mu_i(1+\phi_i) + \phi_i + 2) \right] \frac{\p^2 \mu_i}{\p \eta^2_{1i}} \frac{\p \phi_i}{\p \eta_{2i}} \right.\\
		&-& \frac{(1+\phi_i)^2}{2} \left. \left[ c_i \mu_i  - \psi^{(2)}(\mu_i(1+\phi_i) + \phi_i + 2) \right] \left(\frac{\p \mu_i}{\p \eta_{1i}} \right)^2 \frac{\p \phi_i}{\p \eta_{2i}} \right),\\
M_3 &=& \diag \left( - \frac{(1+ \phi_i)}{2} \left\{ \left[2a_i + (1+\phi_i)c_i\mu_i - (1+\phi_i)\psi^{(2)}(\mu_i(1+\phi_i) + \phi_i + 2) \right] \left(\frac{\p \mu_i}{\p \eta_{1i}} \right)^2 \frac{\p \phi_i}{\p \eta_{2i}} \right. \right. \\
		&+& \left. \left [\psi^{(1)}(\mu_i(1+\phi_i) + \phi_i + 2) - a_i \mu_i] \frac{\p^2 \mu_i}{\p \eta^2_{1i}}\frac{\p \phi_i}{\p \eta_{2i}} \right\} \right),\\
M_4 &=& \diag \left(\frac{1}{2} \left\{[(1+\phi_i)d_i + 2\psi^{(1)}(\mu_i(1+\phi_i) + \phi_i + 2) - 2a_i\mu_i]\frac{\p \mu_i}{\p \eta_{1i}}\left(\frac{\p \phi_i}{\p \eta_{2i}}\right)^2 \right. \right. \\
		&-&\left. \left.  \left(1 + \phi_i\right)[\psi^{(1)}(\mu_i(1+\phi_i) + \phi_i + 2) - a_i \mu_i] \frac{\p \mu_i}{\p \eta_{1i}} \frac{\p^2 \phi_i}{\p \eta^2_{2i}} \right\} \right),\\
M_5 &=& \diag \left( \frac{(1 + \phi_i)}{2} \left[d_i \left(\frac{\p \phi_i}{\p \eta_{2i}}\right)^2\frac{\p \mu_i}{\p \eta_{1i}} + \psi^{(1)}(\mu_i(1+\phi_i) + \phi_i + 2) -a_i\mu_i\right]\frac{\p \mu_i}{\p \eta_{1i}}\frac{\p^2 \phi_i}{\p \eta^2_{2i}} \right),\\
M_6 &=& \diag \left(\frac{1}{2} \left[e_i\left(\frac{\p \phi_i}{\p \eta_{2i}}\right)^3 - b_i\frac{\p \phi_i}{\p \eta_{2i}}\frac{\p^2 \phi_i}{\p \eta_{2i}^2} \right] \right).
\end{eqnarray*}
Considering the matrices defined above, we have that $m_{ij}$ represent the $i$th element of $M_j.$ So, we have
\begin{eqnarray*}
\kappa_{rs}^{(u)} - \frac{1}{2} \kappa_{rsu} &=& \sum_{i=1}^n m_{1i} x_{ir}x_{is}x_{iu},\\
\kappa_{Rs}^{(u)} - \frac{1}{2} \kappa_{Rsu} &=& \sum_{i=1}^n m_{2i} x_{is}x_{iu}z_{iR},\\
\kappa_{rS}^{(u)} - \frac{1}{2} \kappa_{rSu} &=& \sum_{i=1}^n m_{2i} x_{ir}x_{iu}z_{iS},\\
\kappa_{rs}^{(U)} - \frac{1}{2} \kappa_{rsU} &=& \sum_{i=1}^n m_{3i} x_{ir}x_{is}z_{iU},\\
\kappa_{RS}^{(u)} - \frac{1}{2} \kappa_{RSu} &=& \sum_{i=1}^n m_{4i} x_{iu}z_{iR}z_{iS},\\
\kappa_{Rs}^{(U)} - \frac{1}{2} \kappa_{RsU} &=& \sum_{i=1}^n m_{5i} x_{is}z_{iU}z_{iR},\\
\kappa_{rS}^{(U)} - \frac{1}{2} \kappa_{rSU} &=& \sum_{i=1}^n m_{5i} x_{ir}z_{iS}z_{iU},\\
\kappa_{RS}^{(U)} - \frac{1}{2} \kappa_{RSU} &=& \sum_{i=1}^n m_{6i} z_{iR}z_{iS}z_{iU},\\
\end{eqnarray*}
We now obtain the terms from (\ref{vcs}), presenting in detail the algebra to obtain the first term of this expression. To calculate the other terms, we follow the same logic.
\begin{eqnarray*}
\sum_{r,s,u} \kappa^{ar}\kappa^{su} \left\{\kappa_{rs}^{(u)} - \frac{1}{2} \kappa_{rsu} \right\} &=& \sum_{r,s,u} \left(\kappa^{ar} \kappa^{su} \sum_{i=1}^n m_{1i}x_{ir}x_{is}x_{iu} \right) \\
						&=& \sum_{i=1}^n m_{1i} \left(\sum_{r} x_{ir} \kappa^{ar} \right) \left(\sum_{s,u} x_{ir} \kappa^{su} x_{iu} \right)\\
						&=& \sum_{i=1}^n m_{1i} \left(\sum_{r} x_{ir} \kappa^{ar} \right) \bm{\delta}_i^\top (X \kappa^{\bm{\beta \beta}} X^\top)\bm{\delta}_i\\
						&=& \bm{\delta}_a^\top \sum_{i=1}^n \kappa^{a \bm{\beta}}X^\top \bm{\delta}_i m_{1i} \bm{\delta}_i^\top (X \kappa^{\bm{\beta \beta}} X^\top)						 \bm{\delta}_i\\
						&=& \bm{\delta}_a^\top \kappa^{a \bm{\beta}} X^\top M_1 P_{\bm{\beta \beta}},
\end{eqnarray*}
being $\kappa^{a \bm{\beta}}$ the matrix $\kappa^{\bm{\beta \beta}}$ if $a=1,\ldots,p$ and $\kappa^{\bm{\nu \beta}}$ if $a=p+1,\ldots,q,$ $\bm{\delta}_a$ ($\bm{\delta}_i$) an $n \times 1$ vector with a one in the $a$th ($i$th) position. Also, the vector $P_{\bm{\beta \beta}}$ is presented in Section \ref{sec3}. Likewise, we have the remaining quantities expressed by

\begin{eqnarray*}
\sum_{R,s,u} \kappa^{aR}\kappa^{su} \left\{\kappa_{Rs}^{(u)} - \frac{1}{2} \kappa_{Rsu} \right\} &=& \bm{\delta}_a^\top \kappa^{a \bm{\nu}} Z^\top M_2 P_{\bm{\beta \beta}},\\
\sum_{r,S,u} \kappa^{ar}\kappa^{Su} \left\{\kappa_{rS}^{(u)} - \frac{1}{2} \kappa_{rSu} \right\} &=& \bm{\delta}_a^\top \kappa^{a \bm{\beta}} X^\top M_2 P_{\bm{\beta \nu}},\\
\sum_{r,s,U} \kappa^{ar}\kappa^{sU} \left\{\kappa_{rs}^{(U)} - \frac{1}{2} \kappa_{rsU} \right\} &=& \bm{\delta}_a^\top \kappa^{a \bm{\beta}} X^\top M_3 P_{\bm{\beta \nu}},\\
\sum_{R,S,u} \kappa^{aR}\kappa^{Su} \left\{\kappa_{RS}^{(u)} - \frac{1}{2} \kappa_{RSu} \right\} &=& \bm{\delta}_a^\top \kappa^{a \bm{\nu}} Z^\top M_4 P_{\bm{\beta \nu}},\\
\sum_{R,s,U} \kappa^{aR}\kappa^{sU} \left\{\kappa_{Rs}^{(U)} - \frac{1}{2} \kappa_{RsU} \right\} &=& \bm{\delta}_a^\top \kappa^{a \bm{\nu}} Z^\top M_5 P_{\bm{\beta \nu}},\\
\sum_{r,S,U} \kappa^{ar}\kappa^{SU} \left\{\kappa_{rS}^{(U)} - \frac{1}{2} \kappa_{rSU} \right\} &=& \bm{\delta}_a^\top \kappa^{a \bm{\beta}} X^\top M_5 P_{\bm{\nu \nu}},\\
\sum_{R,S,U} \kappa^{aR}\kappa^{SU} \left\{\kappa_{RS}^{(U)} - \frac{1}{2} \kappa_{RSU} \right\} &=& \bm{\delta}_a^\top \kappa^{a \bm{\nu}} Z^\top M_6 P_{\bm{\nu \nu}},
\end{eqnarray*}
where $\kappa^{a \bm{\nu}}$ is the matrix $\kappa^{\bm{\beta \nu}}$ if $a=1,\ldots,p$ and $\kappa^{\bm{\nu \nu}}$ if $a=p+1,\ldots,q$ and the vectors $P_{\bm{\beta \nu}} $ and $P_{\bm{\nu \nu}}$ were presented in Section \ref{sec3}.

\end{document}